\documentclass{ws-procs9x6}

\begin{document}

\title{Energy flow between jets in the $k_t$ algorithm.
%\footnote{\uppercase{T}his work is supported by etc, etc.}
}

\author{M.~Dasgupta 
\footnote{\uppercase{W}ork done in collaboration with \uppercase{A}ndrea \uppercase{B}anfi.
%supported by grant 2-4570.5 of the \uppercase{S}wiss 
%\uppercase{N}ational \uppercase{S}cience \uppercase{F}oundation.
}}

\address{School of Physics and Astronomy,\\
Schuster Building, Brunswick Street, \\ 
Manchester, M13 9PL,\\
United Kingdom. \\ 
E-mail: Mrinal.Dasgupta@manchester.ac.uk}

\maketitle

\abstracts{We consider the impact of the $k_t$ algorithm on energy flow into 
gaps between jets in any QCD hard process. 
While we confirm the observation 
that the $k_t$ clustering procedure considerably reduces the impact of 
non-global logarithms, we unearth 
yet new sources of logarithmic enhancement, that stem from using the $k_t$ algorithm to define the final state. 
We comment on the nature of the logarithms we find 
and discuss their all-orders treatment.}

\section{Introduction}
The transverse energy ($E_t$) flow into gaps between hard 
jets is an observable 
that can offer important insights into different aspects of QCD. 
This includes information on the strong coupling, understanding of ``all-order'' behaviour as manifested in resummed predictions, non-perturbative power corrections and the underlying event at hadron colliders.

In order however to obtain information as accurately as possible from 
such an observable, one might expect that the minimum requirement is 
a solid (and correct) perturbative estimate, at least to the accuracy claimed.
Failure to provide a correct estimate leads to attributing a potentially significant chunk 
of the 
model independent perturbative answer to  model dependent pieces such as 
the underlying event or power corrections. How significant such a 
mis-attribution is, will naturally vary on a case-by--case basis but it 
suffices to say that the overall picture emerging from most studies 
of this kind, would be incomplete. This unfortunately is in fact the 
current situation, as described below.
\section{Non global logs and the $k_t$ definition}
We wish to consider the distribution in the $E_t$ flow into a gap 
$\Omega$, $1/\sigma d\sigma/d Q_\Omega$. 
We define the gap transverse energy as 
\begin{equation}
Q_\Omega = \sum_{i} E_{t,i}
\end{equation}
where the sum runs over either hadrons in $\Omega$ or as is more commonly the case, over soft jets in $\Omega$. These are obtained after a jet algorithm has been employed to cluster the final state into jets. This leaves aside from the high $E_t$ hard jets outside $\Omega$, soft jets that can populate the gap region.

The main problem in obtaining a resummed perturbative prediction for this observable is its non-global nature, that has itself been pointed out only relatively recently \cite{DassalNg1,DassalNG2}. 
Thus while for several observables such as many 
event shapes, one can obtain a next-to--leading log prediction by considering a veto on real emissions attached just to the hard emitting partons, this is not the case here even at leading logarithmic accuracy.
For non global observables like the gap energy distribution the leading 
single-logarithmic resummed result can be expressed as 
(we consider first the definition involving a sum over hadrons in the gap)
\begin{equation}
\frac{1}{\sigma} \frac{d \sigma }{d Q_\Omega} \approx \frac{1}{\sigma} \frac{d}{d Q_\Omega}\left( e^{-R\left(Q/Q_\Omega \right)} S \left (Q/Q_\Omega \right)\right ),
\end{equation}
where the factor $e^{-R}$ represents uncanceled virtual emissions attached just to 
the hard jets, integrated over the gap region. 
The factor $S$ is the non-global term, where one has to consider a soft gluon emitted in 
$\Omega$ as being coherently emitted from an arbitrarily complex ensemble 
consisting not merely of the hard jets but additionally any number of soft gluons outside $\Omega$. Like the term $e^{-R}$, the factor $S$ also resums a class of single-logarithms, $\alpha_s^n \ln^n Q/Q_\Omega$. Till date the 
calculation of the non-global term $S$ has only been performed in the large $N_c$ approximation, making the non-global piece the dominant source of perturbative uncertainty at low $Q_\Omega$. 

A partial solution to the problem was proposed by Appleby and Seymour \cite{AS1}, who pointed out that in several experimental studies one actually employs the definition based on summing over soft jets given by $k_t$ clustering. 
They assumed that the factor $e^{-R}$ was left intact by the clustering 
procedure, since it can be considered as exponentiating a result, $R$, 
obtained by considering just a single emission and its virtual counterpart. 
On the other hand the non-global piece involves multiple emission and 
has thus to be recomputed 
for the case of clustering. In particular gluons that fly into the gap can be pulled out of the gap 
by harder gluons outside which reduces the non-global component significantly, but does not eliminate it altogether.
Thus the Appleby Seymour result assumed the form 
\begin{equation}
\label{eq:as}
\frac{1}{\sigma} \frac{d \sigma }{d Q_\Omega} \approx \frac{1}{\sigma} \frac{d}{d Q_\Omega}\left( e^{-R\left(Q/Q_\Omega \right)} S_{\mathrm{kt}} \left (Q/Q_\Omega \right)\right ), 
\end{equation}
where $S_{\mathrm{kt}} $ is the non-global contribution recomputed with $k_t$ clustering. This 
new non-global correction was found to be less than 20 $\%$ of the unclustered result.

\section{Additional real-virtual mismatch induced by $k_t$ clustering}
Now we reconsider the result  \eqref{eq:as}  and show that it is incorrect in the sense that it does not capture all the relevant single-logarithms even leaving aside those suppressed by $1/N_c^2$ \cite{bandaslett}.  

Let us concentrate on the factor $e^{-R}$ where for the simple case of $e^{+}e^{-}$ annihilation 
$R \sim  C_F \alpha_s(Q) \ln Q/Q_{\Omega}$.  This term represents purely 
virtual emissions above the scale $Q_\Omega$, integrated in a phase space corresponding to 
the gap region. Real emissions below the scale $Q_\Omega$ have been assumed to totally cancel 
while those above $Q_\Omega$ are vetoed.

In fact real emissions attached to the hard jets (thus not pertaining to the non-global term) {\emph{do not completely cancel away}}, due to the use of clustering. 
Consider two energy-ordered real emissions $k_1$ and $k_2$ for which 
the probability of independent emission from a hard dipole $ab$, 
can be written as :
\begin{equation}
{\mathcal{P}}^{\mathrm{real,real}} = C_F^2 \alpha_s^2 W_{ab}(k_1) W_{ab}(k_2),
\end{equation}
where the $W_{ab}$ are eikonal emission factors for gluons $k_1$ and $k_2$  
from the 
$ab$ dipole.
Likewise if the emitting (more energetic) gluon $k_1$ is virtual we have the 
one-real one--virtual independent emission probability.

\begin{equation}
{\mathcal{P}}^{\mathrm{real,virtual}} = -C_F^2 \alpha_s^2 W_{ab}(k_1) W_{ab}(k_2).
\end{equation}
For the pure virtual piece $e^{-R}$ to be built up, one assumes these 
contributions to cancel, which is the case without clustering. 

In the case one uses clustering on the final state, consider a situation when 
the softer gluon $k_2$ is in $\Omega$ and $k_1$ outside. 
If the distance between the gluons 
$\left( \Delta \eta_{12} \right )^2 +\left(\Delta \phi_{12} \right )^2 < R^2 $, , where $R$ is the jet radius, 
the gluon $k_2$ is clustered out of the gap and hence in this region the 
double real contribution to the energy distribution in $\Omega$ is zero. 
However the mixed real-virtual term persists in this region, making a finite contribution to the distribution, since $k_2$ cannot be clustered 
away by a virtual gluon outside the gap. Thus instead of a cancellation we are 
left with a contribution that at order $\alpha_s^2$ has the colour factor 
$C_F^2$. This 
is clearly distinct from the non-global term at this order which has colour factor $C_F C_A$ and is not accounted for by expanding $e^{-R}$ either, 
confirming that it is a piece left out previously.

Specialising to the case of $\Omega$ being a rapidity slice of width 
$\Delta \eta \geq R $ we obtain the following additional single-logarithmic 
contribution to the integrated quantity 
$ \int_0^{Q_\Omega} 
\frac{1}{\sigma}\frac{d\sigma}{d E_t} dE_t $:
\begin{equation}
C_2^{\mathrm{primary}} = \frac{16}{\pi} C_F^2 L^2 R^3,
\end{equation}
where $L = \ln Q/Q_\Omega$ and ``primary'' refers to the fact that we have 
attached the gluons only to the primary hard partons produced in the process.
We have confirmed the result above, valid for 
the simple case of $e^{+}e^{-} \to 2$ jets with exact 
fixed-order computations.

\section{All orders contribution}
We have been able to numerically resum the terms we describe above, to all orders for the simple $e^{+}e^{-} \to$ 2 jets and DIS (1+1) jets cases. 
The effect we find is moderate over most of the phenomenological region of 
interest, changing the previous results by a maximum of $30 \%$. 
However for the more complex cases of dijet 
photoproduction and hadron--hadron energy flow variables, further work is needed to estimate this effect at all orders.
In these cases additional insight is also required to understand the potential role of superleading logarithms \cite{FKS}. A satisfactory understanding of the energy flow even to leading logarithmic accuracy, is thus some way off.


\begin{thebibliography}{0}
\bibitem{DassalNg1} M. Dasgupta and G. P. Salam, {\it Phys. Lett.}
{\bf B512}, 323 (2001).

\bibitem{DassalNG2} M. Dasgupta and G. P. Salam, {\it JHEP}  
{\bf 0203}, 017 (2002).

\bibitem{AS1} R.B. Appleby and M. H. Seymour, {\it JHEP} {\bf 0212}, 063 (2002).

\bibitem{bandaslett} A. Banfi and M. Dasgupta {\it Phys. Lett.} {\bf B 628}, 49  (2005).

\bibitem{FKS} J. Forshaw, A. Kyrieleis and M. H. Seymour, hep-ph/0604094.



\end{thebibliography}
\end{document}